\documentclass[12pt]{article}

\usepackage[normalem]{ulem}
\usepackage[mathscr]{eucal}
\usepackage[version=3]{mhchem}
\usepackage{stmaryrd}

\usepackage{epsfig,latexsym,amsfonts,amsmath,amsthm,amssymb,amsbsy,multirow,slashed,wasysym,textcomp,dsfont,comment,mathtools,cancel,cite,datetime,appendix,BOONDOX-calo}
\usepackage{tocloft}
\usepackage{bbold}
\usepackage{tikz}
\usetikzlibrary{backgrounds}
\usetikzlibrary{patterns}
\usetikzlibrary{arrows.meta}
\tikzstyle{bag} = [align=center]
\usetikzlibrary{decorations.pathmorphing}
\def\bea{\begin{eqnarray}}
\def\eea{\end{eqnarray}}

\usepackage{hyperref}
\usepackage{subcaption}

\newcommand{\badat}{\begin{alignedat}}
\newcommand{\eadat}{\end{alignedat}}
\newcommand\scalemath[2]{\scalebox{#1}{\mbox{\ensuremath{\displaystyle #2}}}}
\def\be{\begin{equation}}
\def\ee{\end{equation}}
\def\ba{\begin{aligned}}
\def\ea{\end{aligned}}

\def\p{\partial}

\usepackage{xcolor}

\newcommand{\pink}[1]{\textcolor{\pink}{#1}}

\definecolor{dblue}{rgb}{0.2,0.50,0.80}

\def\bz{{\bar z}}

\def\zb{\bar{z}}

\def\bz{{\bar z}}

\def\pa{{\partial}}

\def\D{{\Delta}}
\def\d{{\delta}}
\def\o{{\omega}}

\def\s{{\sigma}}

\def\oh{{\mathcal O}}
\def\eps{\varepsilon}
\def\kap{\kappa}
\def\mc{\mathcal}

\newcommand{\res}[1]{\underset{#1}{\rm Res}}

\newcommand{\half}{{1\over 2}}

\DeclareFontFamily{OT1}{pzc}{}
\DeclareFontShape{OT1}{pzc}{m}{it}{<-> s * [1.10] pzcmi7t}{}
\DeclareMathAlphabet{\mathpzc}{OT1}{pzc}{m}{it}
\DeclareMathAlphabet{\mathcal}{OMS}{cmsy}{m}{n}

\definecolor{vert}{rgb}{0.1367 0.543 0.1367}

\numberwithin{equation}{section} 

\begin{document}

 \begin{titlepage}
  \thispagestyle{empty}
  \begin{flushright}
  \end{flushright}
  \bigskip
  \begin{center}

        \baselineskip=13pt {\LARGE \scshape{
       Multicollinear Singularities \\[.5em]  in Celestial CFT
       }}
     
      \vskip1cm 

   \centerline{ 
   {Adam Ball}, ${}^{\clubsuit{},\diamondsuit{}}$
   {Yangrui Hu}, ${}^\diamondsuit{}$
    and {Sabrina Pasterski} ${}^\diamondsuit{}$
}

\bigskip\bigskip
 
 \centerline{\em${}^\clubsuit$ 
\it Department of Physics, Brown University, Providence, RI 02912, USA}

\bigskip
 
 \centerline{\em${}^\diamondsuit$ 
\it Perimeter Institute for Theoretical Physics, Waterloo, ON N2L 2Y5, Canada}

\bigskip\bigskip

\end{center}

\begin{abstract}
 \noindent 

The purpose of this paper is to study the holomorphic multicollinear limit of (celestial) amplitudes and use it to further investigate the double residue condition for (hard celestial) amplitudes and the celestial operator product expansion. We first set up the notion of holomorphic multicollinear limits of amplitudes and derive the 3-collinear splitting functions for Yang-Mills theory, Einstein gravity, and massless $\phi^3$ theory. In particular, we find that in $\phi^3$ theory the celestial 3-OPE contains a term with a branch cut. This explicit example confirms that branch cuts can obstruct the double residue condition for hard celestial amplitudes, which is the underlying cause of the celestial Jacobi identities not holding for certain theories. This addresses an ongoing debate in the literature about associativity of the celestial OPEs and concretely demonstrates a new (multi-particle) term in the celestial OPE coming from the multi-particle channel in the amplitudes.

\end{abstract}

\end{titlepage}

\tableofcontents

\section{Introduction}\label{sec:intro}

The Celestial Holography program aims to apply the holographic principle to asymptotically flat spacetimes, proposing that gravitational scattering can be captured by a CFT living on the codimension-two celestial sphere~\cite{Pasterski:2021rjz,Pasterski:2021raf,Raclariu:2021zjz}. In particular, ${\cal S}$-matrix elements in four dimensions are recast as correlation functions of primaries in a 2D CFT via a dimensional reduction of the conformal boundary that puts the scattering states into boost eigenstates. Once we understand the spectrum of operators that capture the bulk scattering states~\cite{Pasterski:2017kqt,kp}, exploiting the operator product expansion in the 2D CFT would amount to bootstrapping amplitudes from their collinear limits.

The main motivation for this program came from the manner in which a boost basis for scattering appears to naturally organize certain asymptotic symmetry enhancements associated with universal soft theorems~\cite{Strominger:2017zoo}. Rather than just Poincar\'e invariance, the leading and subleading soft graviton theorems gave an angle-dependent enhancement that matched the BMS group~\cite{Bondi:1962px,Sachs:1962wk, Sachs:1962zza}.
More surprisingly, the 2D CFT treatment unveiled a full tower of symmetries arising from leading holomorphic collinear limits~\cite{Guevara:2021abz,Strominger:2021mtt}.
In particular, the tower of soft modes, whose collinear residues truncate into finite-dimensional ${\rm SL}(2,\mathbb{R})$ multiplets, forms the wedge $w_{1+\infty}$ symmetry algebra of gravity of which chiral-Poincar\'e fills the first two entries.
In more detail, these symmetry algebras were originally derived from the following steps~\cite{Guevara:2021abz,Strominger:2021mtt}:
\begin{enumerate}
    \item complexify the celestial sphere and take the collinear limit of two gauge bosons (gluons or gravitons),
    \item define soft currents as the residues at integer scaling dimensions $k\le 2$ and extract their modes via a contour integral
    \begin{equation}
        H^{k,s}(z,\bz) ~=~ \res{\D\to k}\,{\cal O}_{\D,s}(z,\bz) ~~,~~ 
        H^{k,s}_m(z) ~=~ \oint\frac{d\bz}{2\pi i}\,\bz^{\frac{k-s-2}{2}+m}\,H^{k,s}(z,\bz)~~,
    \end{equation}
    where assuming the primary descendants vanish would imply the range of $m$ is restricted,
    \item evaluate the soft current algebra via the radial quantization bracket
    \begin{equation}    \Big[H^{k_1,s_1}_m(z),H^{k_2,s_2}_n(w)\Big]_{w}~=~\oint_{w}\frac{dz}{2\pi i}\,H^{k_1,s_1}_m(z)\,H^{k_2,s_2}_n(w)~~.
    \label{equ:bracket}
    \end{equation}
\end{enumerate}
These holographic symmetries were originally derived from just the holomorphic collinear singularity of the 3-point vertex of the gauge bosons coupling to themselves. In minimally coupled theories this is the only contribution to the bracket defined in (\ref{equ:bracket}).

The steps in the construction of \cite{Guevara:2021abz} straightforwardly generalize to generic 3-point couplings in~\cite{Himwich:2021dau}, although once you change what two same-helicity gauge bosons can couple to, you might expect to deform the algebra.  Such is the case for higher derivative interactions in non-minimally coupled theories. However, more troublesomely, it was shown in \cite{Mago:2021wje} that the resulting na\"ive current algebras generically fail to satisfy the Jacobi identity. 
This came as a surprise since usually in 2D CFT contexts the Jacobi identity follows automatically from the associativity of the modes of the currents. This tension was mostly resolved in \cite{Ball:2022bgg} where it was shown that the non-commutativity of soft limits means that having multiple soft insertions is ill-defined without some further prescription.\footnote{A formulation of CCFT using only integer dimensions~\cite{Freidel:2022skz,Cotler:2023qwh} necessarily involves some further prescription.} This avoids a paradox, but is still somewhat concerning as it seemingly limits the applicability of some of the rich symmetries that are a hallmark of CCFT. In any case, the subset of effective field theories (EFTs) that satisfy Jacobi can be viewed as a constraint submanifold in the space of couplings. In \cite{Ren:2022sws} it was shown for a large family of EFTs that precisely the same coupling constraints follow from a simpler calculation in momentum space, with the Jacobi identity on celestial soft currents replaced by a ``double residue condition" on hard momentum space amplitudes. Namely, the double residue condition is
\be \label{eq:dblmom} 0 ~\stackrel{?}{=}~ \left( \res{z_2\shortrightarrow z_3} \, \res{z_1\shortrightarrow z_2} - \res{z_1\shortrightarrow z_3} \, \res{z_2\shortrightarrow z_3} + \res{z_2\shortrightarrow z_3} \, \res{z_1\shortrightarrow z_3} \right)  \mathcal{A}_n~~. \ee
It can be applied to hard celestial amplitudes just as well as hard momentum space amplitudes, and the authors of \cite{Ren:2022sws} suggested that its failure for the former might be due to non-associativity of the celestial OPE. This would be a bit of an existential crisis for the CCFT construction. In this paper we offer the far more benign resolution of branch cuts on the complexified celestial sphere, which can obstruct the contour pulling argument for the double residue condition. Such branch cuts are ubiquitous in bona fide CFTs (e.g. the Ising model and the free boson), and pose no problems for locality as long as the holomorphic and anti-holomorphic branch cuts cancel in Euclidean signature.

Our results extend those of \cite{Ball:2022bgg}, which demonstrated generally the equivalence between the Jacobi identity on celestial soft currents, the double residue condition on hard celestial amplitudes, the double residue condition on hard momentum space amplitudes, and a certain simple condition on four-point momentum space amplitudes. The mechanisms of failure were also mostly elucidated, with failure of the Jacobi identity for celestial soft currents being due to the non-commutativity of soft limits, and failure of the double residue condition on hard momentum space amplitudes relying on poles from multi-particle factorization channels. However, the mechanism of failure for the double residue condition on hard {\it celestial} amplitudes was yet to be worked out. This motivates us to dive into the multicollinear singularities in celestial amplitudes and observe how they spoil the double residue condition. As we will show in this paper, branch cuts arise essentially from Mellin transforming multi-particle propagators. They even appear in the celestial OPEs, obviating the need for non-associativity.

Indeed, if we trust the fact that the 2D correlators and Hilbert space are coming from the bulk by dimensional reduction of the conformal boundary, then we can argue that associativity of the celestial OPE is inherited from that of the bulk. This gives us an interesting perspective on the concern about loss of celestial symmetries in generic EFTs. From the 4D associativity, we expect that the phase space realization of the celestial symmetries will obey the Jacobi identity.  These generators were constructed to match the undeformed celestial symmetries in~\cite{Freidel:2021ytz,Hu:2022txx,Freidel:2023gue,Hu:2023geb} and, when viewed as only acting on the $in$ or $out$ state, are symmetries of the free theory. It would be interesting to understand the connection between these being symmetries of the ${\cal S}$-matrix and the breakdown in \cite{Mago:2021wje,Ren:2022sws}, though we leave this to future work. The phase space realization nonetheless provides some insight into the non-locality on the complexified celestial sphere we encounter here. Namely, the branch cut in the celestial OPE is naturally associated with a term coming from two-particle operators in the celestial OPE.

\vspace{1em}

In this paper we study holomorphic multicollinear limits of $hard$ amplitudes (both in momentum space and the celestial boost basis).  
The main results are as follows.
\begin{itemize}
    \item We introduce the notion of a holomorphic multicollinear limit. This is distinct from the usual multicollinear limit studied in the amplitudes community within Yang-Mills (for example, see~\cite{Birthwright:2005ak}) where they take both spinors $|\lambda_i\rangle$ and $|\tilde{\lambda}_i]$ collinear whereas we take only $|\lambda_i\rangle$ collinear and leave $|\tilde{\lambda}_i]$ generic. In Yang-Mills these limits give similar results but in general they can be quite different.
    \item We compute the holomorphic 3-collinear all-plus splitting functions for color-ordered Yang-Mills, pure gravity, and $\phi^3$ theory. In particular, for $\phi^3$ theory, the splitting function contains a $1/s_{123} = 1/(p_1+p_2+p_3)^2$ pole, and the leading terms of the celestial 3-OPE are computed which contain a term with a branch cut.  
    This explicit example confirms that branch cuts can obstruct the double residue condition for hard celestial amplitudes.
    \item Finally, we use this result to deduce a new term in celestial two-particle OPE for $\phi^3$ theory.
\end{itemize}

\vspace{1em}
This paper is organized as follows. In section~\ref{sec:multicollinear} we start by introducing the holomorphic multicollinear limit in momentum space and deriving the 3-collinear splitting function for a generic helicity configuration. Mellin transforming this splitting function yields the celestial 3-OPE. For theories that satisfy Jacobi, such as Yang-Mills and Einstein gravity, we show that the $1/s_{123}$ pole cancels out of the splitting function, strengthening a result of~\cite{Ball:2022bgg}. Section~\ref{sec:phi3} is devoted to the specific example of $\phi^3$ theory, where we are able to analytically compute the leading terms of the 3-OPE in a certain limit and use this to demonstrate that it contains a term with a branch cut. From this term we deduce the existence of a new term in the celestial OPE between {\it two} particles in section~\ref{sec:conclusion}.

\paragraph{Conventions}
Before proceeding, let us set some conventions we will use for our computations herein. We parametrize massless momenta as
\be \label{eq:momparam} p^\mu ~=~ \epsilon \, \omega (1+z\zb, z+\zb, -i(z-\zb), 1-z\zb) \ee
with $\epsilon=\pm 1$ indicating states that are outgoing or incoming and $\omega>0$. We nominally use the Minkowski metric $\eta_{\mu\nu} = {\rm diag}(-,+,+,+)$, but we will almost always intend $z, \zb$ as independent complex variables. The corresponding spinors are
\be |p\rangle ~=~ \epsilon \sqrt{2\omega} \begin{pmatrix} 1 \\ z \end{pmatrix}, \quad |p] ~=~ \sqrt{2\omega} \begin{pmatrix} 1 \\ \zb \end{pmatrix} ~~.\ee
The inner product between two momenta $p_1, p_2$ in this parametrization is
\be p_1 \cdot p_2 ~=~ -2\epsilon_1 \epsilon_2 \omega_1 \omega_2 z_{12} \zb_{12} ~~.\ee
The corresponding spinor inner products are
\be \langle 12\rangle ~=~ 2\epsilon_1\epsilon_2 \sqrt{\omega_1\omega_2} \, z_{12} ~~,~~  [12] ~=~ 2\sqrt{\omega_1\omega_2} \, \zb_{12} ~~,\ee
and in particular $s_{12}=(p_1+p_2)^2=2p_1 \cdot p_2=-\langle 12\rangle [12]$. Our convention for 3-point amplitudes is
\be \mc{A}_3(1^{s_1}, 2^{s_2}, 3^{s_3}) ~=~ \Bigg\{ \begin{array}{cc} \kap_{123} [12]^{s_1+s_2-s_3} [23]^{s_2+s_3-s_1} [31]^{s_3+s_1-s_2} & {\rm if} \quad s_1+s_2+s_3 \ge 0 \\ \kap_{123} \langle 12\rangle^{s_3-s_1-s_2} \langle 23\rangle^{s_1-s_2-s_3} \langle 31\rangle^{s_2-s_3-s_1} & {\rm if} \quad s_1+s_2+s_3 \le 0 \end{array}~~. \ee

\section{The Holomorphic Multicollinear Limit and Multi-OPE}\label{sec:multicollinear}

The (holomorphic) celestial OPE was derived in \cite{Fan:2019emx, Pate:2019lpp} by Mellin transforming the (holomorphic) collinear limit.\footnote{In the usual, or ``true", collinear limit one takes two massless momenta to be nearly parallel. In the holomorphic collinear limit one analytically continues away from Lorentzian signature and takes two angle bracket spinors to be nearly parallel, with the corresponding square bracket spinors held fixed.} Similarly in this section we introduce the holomorphic celestial multi-OPE, derived by Mellin transforming the holomorphic multicollinear limit. Multicollinear limits have been studied in the amplitudes literature \cite{Birthwright:2005ak}, and more recently have been applied to celestial amplitudes \cite{Ebert:2020nqf}. However the \textit{holomorphic} multicollinear limit seems to be absent from the literature. As we will see, it differs in general from the true multicollinear limit and is particularly useful in the celestial context. In the following we exclusively study holomorphic limits, but may sometimes omit the word ``holomorphic" for brevity.

\subsection{Momentum space holomorphic multicollinear limit}\label{sec:mom-space-multicollinear}

In this subsection we study the holomorphic 3-collinear splitting function ${\rm Split}[1^{s_1}2^{s_2}3^{s_3}\to J^{s_J}]$ with arbitrary integer helicities $s_1$, $s_2$, and $s_3$. To reduce clutter we assume that our massless momenta $p_1, p_2, p_3$ are all future-directed (i.e. $\epsilon_1=\epsilon_2=\epsilon_3=1$). We parametrize the 3-collinear limit as
\be z_1 ~=~ z_3 + \eps~~,~~ z_2 ~=~ z_3 + \eta \, \eps~~, \ee
or equivalently
\be \ba |1\rangle ~& =~ \sqrt{2\omega_1} \left( |\hat 3\rangle + \eps |r\rangle \right) ~~,\\
|2\rangle ~& =~ \sqrt{2\omega_2} \left( |\hat 3\rangle + \eta \, \eps |r\rangle \right) ~~,\\
|3\rangle ~& =~ \sqrt{2\omega_3} \, |\hat 3\rangle~~, \ea \ee
where the reference spinor is
\be |r\rangle ~=~ \begin{pmatrix} 0 \\ 1 \end{pmatrix}~~. \ee
As we take $\eps\to 0$ with other parameters fixed the most divergent possible behavior is $\oh(\eps^{-2})$, coming from diagrams with the propagator $1/s_{123}$ and one of $1/s_{12}$, $1/s_{13}$, or $1/s_{23}$. The three special cases of $\eta=0,1,\infty$ recover the three possible consecutive collinear limits one can take on $z_1, z_2, z_3$, but for generic $\eta$ there is no simple relation to consecutive collinear limits; the 3-collinear limit contains strictly more information. Furthermore, it is not possible to obtain the splitting function as a limit of a 4-point amplitude. Still, we can be very explicit by working in terms of on-shell 3-point amplitudes. For concreteness, let $\mc{A}_n$ denote an $n$-point amplitude. First, consider the types of diagrams that factorize on the propagators $1/s_{12}$ and $1/s_{123}$. Keeping only the $\oh(\eps^{-2})$ terms, we have the contribution\footnote{In the diagrams in (\ref{eq:D12}-\ref{eq:D13}) we use cuts to visualize the factorization.}
\be \label{eq:D12} \ba D_{1,2} &\equiv~   \begin{tikzpicture}[baseline={([yshift=-0.9ex]current bounding box.center)},scale=0.9]
        \draw[thick] (-1.2,0) -- (-0.6,0) -- (-0.2,0);
        \draw[thick] (-0.2+0.4+0.1,0) -- (0.2+0.2+0.1,0) node[above]{\footnotesize{$J$}}  -- (1.2+0.1,0);
        \draw[thick] (-1.5,0) -- (-1.5-1.12583,-0.65) node[left]{$3$};
        \draw[thick] (-1.5,0) -- (-1.5-1.12583,0.65);
        \draw (-1.5-1.12583-0.1,0.65-0.6) node[above]{\footnotesize{$I$}};
        \draw[thick] (1.5+0.4+0.1,0) -- (1.5+0.4+0.1+1.47224,-0.85) node[right]{$4$};
        \draw[thick] (1.5+0.4+0.1,0) -- (1.5+0.4+0.1+1.47224,0.85) node[right]{$n$};
        \draw (1.5+0.4+0.1+1.47224+0.05,0.1) node[right]{\bf{$\vdots$}} ;
        \filldraw[thick,fill=white] (-1.5,0) circle (0.4) node{{$\kappa$}};
        \filldraw[thick,fill=white] (1.5+0.4+0.1,0) circle (0.8) node{};
        \filldraw[pattern=north east lines] (1.5+0.4+0.1,0) circle (0.8) node{};
        \draw (-0.6+0.1,0) node[above]{\footnotesize{$-J$}};
        \draw[thick] (-1.5-1.12583-0.34641-0.1732,0.65+0.2+0.1) -- (-1.5-1.12583-0.34641-1.12583-0.1732,0.65+0.2+0.65+0.1) ;
        \draw[thick] (-1.5-1.12583-0.34641-1.12583-0.1732,0.65+0.2+0.65+0.1+0.1732) -- (-1.5-1.12583-0.34641-1.12583-1.12583-0.1732,0.65+0.2+0.1) node[left]{$2$};
        \draw[thick] (-1.5-1.12583-0.34641-1.12583-0.1732,+0.65+0.2+0.65+0.1) -- (-1.5-1.12583-0.34641-1.12583-1.12583-0.1732,0.65+0.2+0.65+0.65+0.1) node[left]{$1$};
        \filldraw[thick,fill=white] (-1.5-1.12583-0.34641-1.12583-0.1732,0.65+0.2+0.65+0.1) circle (0.4) node{{$\kappa$}};
        \draw (-1.5-1.12583-0.34641-0.5-0.25-0.1732,0.65) node[right]{\footnotesize	{$-I$}};
\end{tikzpicture} \\
& =~ \scalemath{0.9}{\sum_{I, J}} \mc{A}_3(1^{s_1}, 2^{s_2}, -I^{-s_I}) \frac{1}{p_I^2} \mc{A}_3(I^{s_I}, 3^{s_3}, -J^{-s_J}) \frac{1}{p_J^2} \mc{A}_{n-2}(J^{s_J}, \dots, n^{s_n}) \\
& =\, \scalemath{0.9}{\sum_{I,J} \kappa_{1,2,-I} \kappa_{I,3,-J} \frac{[12]^{s_1+s_2+s_I} [2I]^{s_2-s_I-s_1} [I1]^{-s_I+s_1-s_2}}{-\langle 12\rangle [12]} \, \frac{[I3]^{s_I+s_3+s_J} [3J]^{s_3-s_J-s_I} [JI]^{-s_J+s_I-s_3}}{-\langle 12\rangle [12] - \langle 13\rangle [13] - \langle 23\rangle [23]}} \mc{A}_{n-2} \\
& =~ \frac{1}{(1-\eta)\eps^2} \scalemath{.9}{\sum_{I,J}} \frac{\kappa_{1,2,-I} \kappa_{I,3,-J}}{2^{4-s_1-s_2-s_3+s_J}} \\
& \qquad \times \frac{\omega_1^{s_2-s_I-1} \omega_2^{s_1-s_I-1} \omega_3^{s_I-s_J}}{(\omega_1+\omega_2+\omega_3)^{-s_J}} \frac{\zb_{12}^{s_1+s_2-s_I-1} (\omega_1 \zb_{13} + \omega_2 \zb_{23})^{s_3+s_I-s_J}}{\omega_1\omega_2 (1-\eta) \zb_{12} + \omega_1\omega_3 \zb_{13} + \omega_2\omega_3 \eta \zb_{23}} {\cal A}_{n-2} \ea \ee
where $z_{ij} = z_i - z_j$, the sum is over massless particles and helicities such that $s_1+s_2-s_I \ge 0$ and $s_I + s_3 - s_J \ge 0$, and in the 3-point amplitudes we set $\eps=0$ so that $p_I, p_J$ are massless. The helicity constraint is so that the 3-point amplitudes are independent of $z_{ij}$. We have also used the following relations, which hold at $\eps=0$:
\be \omega_I ~=~ \omega_1 + \omega_2~~, ~~ \zb_I ~=~ \frac{\omega_1 \zb_1 + \omega_2 \zb_2}{\omega_1 + \omega_2}~~, \ee
\be \omega_J ~=~ \omega_1 + \omega_2 + \omega_3~~,~~ \zb_J ~=~ \frac{\omega_1 \zb_1 + \omega_2 \zb_2 + \omega_3 \zb_3}{\omega_1 + \omega_2 + \omega_3}~~. \ee
The two other contributing cases are
\be \label{eq:D23} \ba D_{2,3} ~& \equiv~         \begin{tikzpicture}[baseline={([yshift=-0.9ex]current bounding box.center)},scale=0.9]
        \draw[thick] (-1.2,0) -- (-0.6,0) -- (-0.2,0);
        \draw[thick] (-0.2+0.4+0.1,0) -- (0.2+0.2+0.1,0) node[above]{\footnotesize{$J$}}  -- (1.2+0.1,0);
        \draw[thick] (-1.5,0) -- (-1.5-1.12583,-0.65) node[left]{$1$};
        \draw[thick] (-1.5,0) -- (-1.5-1.12583,0.65);
        \draw (-1.5-1.12583-0.1,0.65-0.6) node[above]{\footnotesize{$I$}};
        \draw[thick] (1.5+0.4+0.1,0) -- (1.5+0.4+0.1+1.47224,-0.85) node[right]{$4$};
        \draw[thick] (1.5+0.4+0.1,0) -- (1.5+0.4+0.1+1.47224,0.85) node[right]{$n$};
        \draw (1.5+0.4+0.1+1.47224+0.05,0.1) node[right]{\bf{$\vdots$}} ;
        \filldraw[thick,fill=white] (-1.5,0) circle (0.4) node{{$\kappa$}};
        \filldraw[thick,fill=white] (1.5+0.4+0.1,0) circle (0.8) node{};
        \filldraw[pattern=north east lines] (1.5+0.4+0.1,0) circle (0.8) node{};
        \draw (-0.6+0.1,0) node[above]{\footnotesize{$-J$}};
        \draw[thick] (-1.5-1.12583-0.34641-0.1732,0.65+0.2+0.1) -- (-1.5-1.12583-0.34641-1.12583-0.1732,0.65+0.2+0.65+0.1) ;
        \draw[thick] (-1.5-1.12583-0.34641-1.12583-0.1732,0.65+0.2+0.65+0.1+0.1732) -- (-1.5-1.12583-0.34641-1.12583-1.12583-0.1732,0.65+0.2+0.1) node[left]{$3$};
        \draw[thick] (-1.5-1.12583-0.34641-1.12583-0.1732,+0.65+0.2+0.65+0.1) -- (-1.5-1.12583-0.34641-1.12583-1.12583-0.1732,0.65+0.2+0.65+0.65+0.1) node[left]{$2$};
        \filldraw[thick,fill=white] (-1.5-1.12583-0.34641-1.12583-0.1732,0.65+0.2+0.65+0.1) circle (0.4) node{{$\kappa$}};
        \draw (-1.5-1.12583-0.34641-0.5-0.25-0.1732,0.65) node[right]{\footnotesize	{$-I$}};
\end{tikzpicture} \\
~& =~ \scalemath{.9}{\sum_{I, J}} \mc{A}_3(2^{s_2}, 3^{s_3}, -I^{-s_I}) \frac{1}{p_I^2} \mc{A}_3(I^{s_I}, 1^{s_1}, -J^{-s_J}) \frac{1}{p_J^2} \mc{A}_{n-2}(J^{s_J}, \dots, n^{s_n}) \\
~& =~ \frac{1}{\eta\eps^2} \scalemath{.9}{\sum_{I,J}} \frac{\kappa_{2,3,-I} \kappa_{I,1,-J}}{2^{4-s_1-s_2-s_3+s_J}} \\
& \qquad \times \frac{\omega_1^{s_I-s_J} \omega_2^{s_3-s_I-1} \omega_3^{s_2-s_I-1}}{(\omega_1+\omega_2+\omega_3)^{-s_J}} \frac{\zb_{23}^{s_2+s_3-s_I-1} (-\omega_2 \zb_{12} - \omega_3 \zb_{13})^{s_1+s_I-s_J}}{\omega_1\omega_2(1-\eta)\zb_{12} + \omega_1\omega_3\zb_{13} + \omega_2\omega_3\eta\zb_{23}} {\cal A}_{n-2} \ea \ee
and
\be \label{eq:D13} \ba D_{1,3} ~& \equiv~     \begin{tikzpicture}[baseline={([yshift=-0.9ex]current bounding box.center)},scale=0.9]
        \draw[thick] (-1.2,0) -- (-0.6,0) -- (-0.2,0);
        \draw[thick] (-0.2+0.4+0.1,0) -- (0.2+0.2+0.1,0) node[above]{\footnotesize{$J$}}  -- (1.2+0.1,0);
        \draw[thick] (-1.5,0) -- (-1.5-1.12583,-0.65) node[left]{$2$};
        \draw[thick] (-1.5,0) -- (-1.5-1.12583,0.65);
        \draw (-1.5-1.12583-0.1,0.65-0.6) node[above]{\footnotesize{$I$}};
        \draw[thick] (1.5+0.4+0.1,0) -- (1.5+0.4+0.1+1.47224,-0.85) node[right]{$4$};
        \draw[thick] (1.5+0.4+0.1,0) -- (1.5+0.4+0.1+1.47224,0.85) node[right]{$n$};
        \draw (1.5+0.4+0.1+1.47224+0.05,0.1) node[right]{\bf{$\vdots$}} ;
        \filldraw[thick,fill=white] (-1.5,0) circle (0.4) node{{$\kappa$}};
        \filldraw[thick,fill=white] (1.5+0.4+0.1,0) circle (0.8) node{};
        \filldraw[pattern=north east lines] (1.5+0.4+0.1,0) circle (0.8) node{};
        \draw (-0.6+0.1,0) node[above]{\footnotesize{$-J$}};
        \draw[thick] (-1.5-1.12583-0.34641-0.1732,0.65+0.2+0.1) -- (-1.5-1.12583-0.34641-1.12583-0.1732,0.65+0.2+0.65+0.1) ;
        \draw[thick] (-1.5-1.12583-0.34641-1.12583-0.1732,0.65+0.2+0.65+0.1+0.1732) -- (-1.5-1.12583-0.34641-1.12583-1.12583-0.1732,0.65+0.2+0.1) node[left]{$3$};
        \draw[thick] (-1.5-1.12583-0.34641-1.12583-0.1732,+0.65+0.2+0.65+0.1) -- (-1.5-1.12583-0.34641-1.12583-1.12583-0.1732,0.65+0.2+0.65+0.65+0.1) node[left]{$1$};
        \filldraw[thick,fill=white] (-1.5-1.12583-0.34641-1.12583-0.1732,0.65+0.2+0.65+0.1) circle (0.4) node{{$\kappa$}};
        \draw (-1.5-1.12583-0.34641-0.5-0.25-0.1732,0.65) node[right]{\footnotesize	{$-I$}};
\end{tikzpicture} \\
~& =~ \scalemath{.9}{\sum_{I, J}} \mc{A}_3(1^{s_1}, 3^{s_3}, -I^{-s_I}) \frac{1}{p_I^2} \mc{A}_3(I^{s_I}, 2^{s_2}, -J^{-s_J}) \frac{1}{p_J^2} \mc{A}_{n-2}(J^{s_J}, \dots, n^{s_n}) \\
~& =~ \frac{1}{\eps^2} \scalemath{.9}{\sum_{I,J}} \frac{\kappa_{1,3,-I} \kappa_{I,2,-J}}{2^{4-s_1-s_2-s_3+s_J}} \\
& \qquad \times \frac{\omega_1^{s_3-s_I-1} \omega_2^{s_I-s_J} \omega_3^{s_1-s_I-1}}{(\omega_1+\omega_2+\omega_3)^{-s_J}} \frac{\zb_{13}^{s_1+s_3-s_I-1} (\omega_1 \zb_{12} - \omega_3 \zb_{23})^{s_2+s_I-s_J}}{\omega_1\omega_2(1-\eta)\zb_{12} + \omega_1\omega_3\zb_{13} + \omega_2\omega_3\eta\zb_{23}} {\cal A}_{n-2}~~. \ea \ee
The key thing to notice in the expressions for the $D_{i,j}$ is the shared factor of
\be \frac{1}{p_J^2} ~=~ \frac{1}{(p_1+p_2+p_3)^2} ~=~ \frac{-1/4\eps}{\omega_1\omega_2 (1-\eta) \zb_{12} + \omega_1\omega_3 \zb_{13} + \omega_2\omega_3 \eta \zb_{23}} ~~.\ee
This is the only inhomogeneous term in $\eta$. The full multicollinear limit is given by the sum of these three contributions,
\be \label{eq:splitting} \ba \mc{A}_n ~& =~ D_{1,2} + D_{2,3} + D_{1,3} + \oh(\eps^{-1}) \\
~& =~ \sum_J {\rm Split}[1^{s_1}2^{s_2}3^{s_3}\to J^{s_J}] \, \mc{A}_{n-2}(J^{s_J}, \dots, n^{s_n}) + \oh(\eps^{-1}) ~~.\ea \ee
We define the splitting function as the order $\eps^{-2}$ coefficient of $\mc{A}_{n-2}(J^{s_J}, \dots, n^{s_n})$. Note that it applies equally well to the amplitude $\mc{A}_n$ and the momentum-stripped amplitude $A_n$, where $\mc{A}_n = A_n \delta^{(4)}(\sum_i p_i)$.

\subsection{Celestial multi-OPE}\label{sec:celestial-mulope}

Upon transforming to get a celestial amplitude, the multicollinear splitting function turns into an object that we call the celestial multi-OPE. The derivation is a direct generalization of that for the celestial OPE in \cite{Fan:2019emx, Pate:2019lpp}. We present here only the result for the 3-OPE. The setup is
\be \tilde{\mc A}_n ~=~ \int\dots \int_0^\infty d\omega_1 \int_0^\infty d\omega_2 \int_0^\infty d\omega_3 \, \omega_1^{\Delta_1-1} \omega_2^{\Delta_2-1} \omega_3^{\Delta_3-1} \mc{A}_n~~. \ee
The ellipsis indicates the appropriate integral transforms for the other legs of $\mc{A}_n$, which may or may not be massless. We assume there are at least four such integrals that can be used to absorb the momentum-conserving delta function in $\mc{A}_n = A_n \delta^{(4)}(\sum_i p_i)$. This will give some smooth factor, which we call $F$, that will come along for the ride in the derivation. The integral of interest is then 
\be I ~\equiv~ \int_0^\infty d\omega_1 \int_0^\infty d\omega_2 \int_0^\infty d\omega_3 \, \omega_1^{\Delta_1-1} \omega_2^{\Delta_2-1} \omega_3^{\Delta_3-1} A_n F ~~.\ee
We take $\eps$ small and approximate $A_n$ by its leading behavior controlled by the splitting function (and set $\eps=0$ in $F$).\footnote{There is a subtlety here. The multicollinear limit applies at any fixed $\omega_1, \omega_2, \omega_3$, but here we are integrating over them, so there will always be part of the integration range where, say, $(p_1+p_2)^2$ is not small. This means that the subleading terms in the $\eps$ series expansion of the integrand cannot be neglected. We will neglect them anyway, following the lead of \cite{Fan:2019emx, Pate:2019lpp} for the derivation of the usual (singular part of the) celestial 2-OPE where the same subtlety arises. We proceed with the understanding that this means our multi-OPE misses certain terms. Still, our multi-OPEs form a useful hierarchy where the $(n+1)$-OPE knows more than the $n$-OPE, and in section \ref{sec:conclusion} we will use our 3-OPE to deduce a previously unknown term in the 2-OPE.} A useful change of variables is
\begin{equation}
    \begin{split}
        \omega ~=~ \omega_1+\omega_2+\omega_3~~,~~ &s ~=~ \frac{\omega_1}{\omega_1+\omega_2+\omega_3}~~, \quad t ~=~ \frac{\omega_2}{\omega_1+\omega_2+\omega_3}~~,\\
        &d\omega_1 d\omega_2 d\omega_3 ~=~ d\omega ds dt \, \omega^2 ~~.
    \end{split}
\end{equation}
We can see from the explicit form of the $D_{i,j}$ above that the splitting function is homogeneous in $\omega$, scaling as
\be {\rm Split}[1^{s_1}2^{s_2}3^{s_3}\to J^{s_J}] ~\sim~ \omega^{s_1+s_2+s_3-s_J-4}~~. \ee
Let us define the $\omega$-independent quantity
\be \widehat{\rm Split}[1^{s_1}2^{s_2}3^{s_3}\to J^{s_J}] ~\equiv~ \omega^{-(s_1+s_2+s_3-s_J-4)} {\rm Split}[1^{s_1}2^{s_2}3^{s_3}\to J^{s_J}]~~. \ee
We then have
\be \ba I ~& =~ \sum_J \int_0^1 ds \int_0^{1-s} dt \, s^{\Delta_1-1} t^{\Delta_2-1} (1-s-t)^{\Delta_3-1} \widehat{\rm Split}[1^{s_1}2^{s_2}3^{s_3}\to J^{s_J}] \\
& \qquad \times \int_0^\infty d\omega \, \omega^{\Delta_1+\Delta_2+\Delta_3+s_1+s_2+s_3-s_J-5} A_{n-2}\, F \quad + \quad \oh(\eps^{-1})~~. \ea \ee
The $d\omega$ integral takes the form of a Mellin transform, and depends on $s, t$ only through $\zb_J = \zb_3 + s \zb_{13} + t \zb_{23}$. Plugging back into the full celestial amplitude $\tilde{\mc A}_n$, we have
\be \ba \tilde{\mc A}_n(& \Delta_1, s_1, z_1, \zb_1; \Delta_2, s_2, z_2, \zb_2; \Delta_3, s_3, z_3, \zb_3; \dots)  \\
~=&~ \sum_J \int_0^1 ds \int_0^{1-s} dt \, s^{\Delta_1-1} t^{\Delta_2-1} (1-s-t)^{\Delta_3-1} \widehat{\rm Split}[1^{s_1}2^{s_2}3^{s_3}\to J^{s_J}] \\
& \times \tilde{\mc A}_{n-2}(\Delta_1+\Delta_2+\Delta_3+s_1+s_2+s_3-s_J-4, s_J, z_3, \zb_3 + s \zb_{13} + t \zb_{23}; \dots) + \oh(\eps^{-1})~~. \ea \ee
This formula expresses the leading multicollinear behavior of an $n$-point celestial amplitude in terms of an $(n-2)$-point celestial amplitude. To get a completely factorized form, we can Taylor expand in $s, t$ to get
\be \ba \tilde{\mc A}_n(&\Delta_1, s_1, z_1, \zb_1; \Delta_2, s_2, z_2, \zb_2; \Delta_3, s_3, z_3, \zb_3; \dots)  \\
~=&~ \sum_J \sum_{m=0}^\infty \int_0^1 ds \int_0^{1-s} dt \, s^{\Delta_1-1} t^{\Delta_2-1} (1-s-t)^{\Delta_3-1} \widehat{\rm Split}[1^{s_1}2^{s_2}3^{s_3}\to J^{s_J}] (s\zb_{13} + t\zb_{23})^m \\
& \quad \times \frac{1}{m!} \p_{\zb_3}^m \tilde{\mc A}_{n-2}(\Delta_1+\Delta_2+\Delta_3+s_1+s_2+s_3-s_J-4, s_J, z_3, \zb_3; \dots) + \oh(\eps^{-1})~~. \ea \ee
Defining
\be \mc{I}_m ~\equiv~ \int_0^1 ds \int_0^{1-s} dt \, s^{\Delta_1-1} t^{\Delta_2-1} (1-s-t)^{\Delta_3-1} \widehat{\rm Split}[1^{s_1}2^{s_2}3^{s_3}\to J^{s_J}] (s\zb_{13} + t\zb_{23})^m~~, \ee
we can rewrite this more compactly as
\be \ba \tilde{\mc A}_n(&\Delta_1, s_1, z_1, \zb_1; \Delta_2, s_2, z_2, \zb_2; \Delta_3, s_3, z_3, \zb_3; \dots)  \\
~=&~ \sum_J \sum_{m=0}^\infty \frac{\mc{I}_m}{m!} \p_{\zb_3}^m \tilde{\mc A}_{n-2}(\Delta_1+\Delta_2+\Delta_3+s_1+s_2+s_3-s_J-4, s_J, z_3, \zb_3; \dots) + \oh(\eps^{-1})~~. \ea \ee
This is the general formula for the (maximally singular in $\eps$) 3-OPE coefficients. Keep in mind ${\cal I}_m$ is $\oh(\eps^{-2})$. In the more familiar $2$-collinear case the OPE coefficients are beta functions depending on the conformal weights, but the 3-OPE coefficients $\mc{I}_m$ here are qualitatively different in that they also have $z_i$- and $\zb_i$-dependence. In particular they depend on the parameter $\eta=z_{23}/z_{13}$ defined above. The dependence of $\mc{I}_m$ on $\eta$ will be our main focus hereafter.

\subsection{Simplification when Jacobi is satisfied}

In theories whose celestial currents satisfy Jacobi there will always be a cancellation of the $1/s_{123}$ propagator in the 3-collinear splitting function, and the result is just an $\eta$-weighted sum over the different consecutive holomorphic collinear limits. We show in appendix \ref{app:dblres} that the double residue condition can be reexpressed in terms of $\eta$ as
\be 0 ~\stackrel{?}{=}~ \left( -\res{\eta\to 1} - \res{\eta\to 0} + \lim_{\eta\to\infty} \eta \right) {\rm Split}[1^{s_1}2^{s_2}3^{s_3}\to J^{s_J}] ~~.\ee
We can see from the explicit general form of the splitting function that it can only have simple poles in $\eta$, and they can only lie at $\eta=0,1,\infty$ and $\eta=\eta_*$, where $s_{123}|_{\eta=\eta_*} = 0$. Contour pulling then shows that
\be \left( -\res{\eta\to 1} - \res{\eta\to 0} + \lim_{\eta\to\infty} \eta \right) {\rm Split}[1^{s_1}2^{s_2}3^{s_3}\to J^{s_J}] ~=~ \res{\eta\to\eta_*} \, {\rm Split}[1^{s_1}2^{s_2}3^{s_3}\to J^{s_J}] ~~.\ee
So the entire double residue condition boils down to
\be 0 ~\stackrel{?}{=}~ \res{\eta\to\eta_*} \, {\rm Split}[1^{s_1}2^{s_2}3^{s_3}\to J^{s_J}] ~~.\ee
If this vanishes then the splitting function, which is rational in $\eta$, must not actually have a pole at $\eta_*$, meaning the $1/s_{123}$ propagator has cancelled with part of the numerator. We now explore how this plays out for Yang-Mills and gravity.

\paragraph{Color-ordered Yang-Mills}

Let us compute ${\rm Split}[1^+2^+3^+\to P^+]$ for color-ordered Yang-Mills by plugging the spin values into \eqref{eq:splitting}. One finds 
\be \ba {\rm Split}[1^+2^+3^+\to P^+] 
~& =~ \frac{\kappa_{++-}^2}{4\eps^2} \frac{(\omega_1+\omega_2+\omega_3) \left( \frac{\omega_1 \zb_{13} + \omega_2 \zb_{23}}{\omega_1\omega_2 (1-\eta)} + \frac{\omega_2 \zb_{12} + \omega_3 \zb_{13}}{\omega_2\omega_3\eta}\right)}{\omega_1\omega_2 (1-\eta) \zb_{12} + \omega_1\omega_3 \zb_{13} + \omega_2\omega_3\eta \zb_{23}} \\
~& =~ \frac{\kappa_{++-}^2}{4\eps^2} \frac{\omega_1+\omega_2+\omega_3}{\omega_1\omega_2\omega_3 \eta(1-\eta)} ~~.\ea \ee
The cancellation that occurred here is closely related to the cancellation leading to the vanishing 4-point amplitude ${\mathcal A}_{+++-} = 0$. The corresponding 3-OPE is
\begin{equation}
    \begin{aligned}
    &{\cal O}_{\D_1}^+(z_1,\bz_1) {\cal O}_{\D_2}^+(z_2,\bz_2) {\cal O}_{\D_3}^+(z_3,\bz_3) ~=~ \frac{\kappa_{++-}^2}{4z_{12}z_{23}}\\
    &\sum_{n=0}^{\infty}\,\sum_{m=0}^n\,B(\D_1-1+m,\D_2-1+n-m,\D_3-1)\, \frac{\bz^m_{13}\,\bz^{n-m}_{23}}{m!(n-m)!}\, \pa_{\zb_3}^n {\cal O}^+_{\D_1+\D_2+\D_3-2}(z_3,\bz_3)
    \end{aligned}
\end{equation}
where $B(x,y,z) = \frac{\Gamma(x) \Gamma(y) \Gamma(z)}{\Gamma(x+y+z)}$ is the multivariate beta function. Note that the $\eta$-dependence of this 3-OPE is single-valued. This result has already appeared in \cite{Ebert:2020nqf} in the context of true multicollinear limits in CCFT. For Yang-Mills, the true and holomorphic multicollinear limits of three positive-helicity gluons have the same leading term. Similar results hold for ${\rm Split}[1^+2^+3^- \to P^-]$ and permutations thereof.

\paragraph{Pure Gravity}
We now compute ${\rm Split}[1^{++}2^{++}3^{++}\to P^{++}]$ for pure Einstein gravity, in a convention where $\kappa_{++,++,--} = \frac{\kappa}{2}$. We find
\be \ba {\rm Split}[& 1^{++} 2^{++}3^{++}\to P^{++}] \\
~& =~ \frac{\kappa^2}{4\eps^2} \frac{(\omega_1+\omega_2+\omega_3)^2 \left( \zb_{12} \frac{(\omega_1 \zb_{13} + \omega_2 \zb_{23})^2}{\omega_1\omega_2 (1-\eta)} + \zb_{23} \frac{(\omega_2 \zb_{12} + \omega_3 \zb_{13})^2}{\omega_2\omega_3\eta} + 
\zb_{13} \frac{(\omega_1 \zb_{12} - \omega_3 \zb_{23})^2}{\omega_1\omega_3} \right)}{\omega_1\omega_2 (1-\eta) \zb_{12} + \omega_1\omega_3 \zb_{13} + \omega_2\omega_3\eta \zb_{23}} \\
~& =~ \frac{\kappa^2}{4\eps^2} \frac{(\omega_1+\omega_2+\omega_3)^2}{\omega_1\omega_2\omega_3} \left( \omega_1\frac{\zb_{12}\zb_{13}}{1-\eta} + \omega_2\frac{\zb_{12}\zb_{23}}{\eta(1-\eta)} + \omega_3\frac{\zb_{13}\zb_{23}}{\eta} \right)~~. \ea \ee
The $1/s_{123}$ propagator has cancelled, and the result is effectively an $\eta$-weighted sum over the three possible consecutive collinear limits. Performing the Mellin transform, we have
\begin{equation}\resizebox{0.9\textwidth}{!}{$%
    \begin{aligned}
        &{\cal O}^{++}_{\D_1}(z_1,\bz_1) {\cal O}^{++}_{\D_2}(z_2,\bz_2) {\cal O}^{++}_{\D_3}(z_3,\bz_3) 
        ~=~ \left(\frac{\kappa}{2}\right)^2\,\sum_{n=0}^{\infty}\sum_{m=0}^n\,\frac{\bz^m_{13}\,\bz^{n-m}_{23}}{m!(n-m)!}\pa_{\zb_3}^n{\cal O}^{++}_{\D_1+\D_2+\D_3}(z_3,\bz_3)\\
        &\Bigg[\,B(\D_1+m,\D_2-1+n-m,\D_3-1)\,\frac{\bz_{12}\bz_{13}}{z_{12}z_{13}} ~+~   B(\D_1-1+m,\D_2+n-m,\D_3-1)\,\frac{\bz_{12}\bz_{23}}{z_{12}z_{23}} \\
        &\qquad\qquad\qquad\qquad\qquad\qquad\qquad\qquad ~+~ B(\D_1-1+m,\D_2-1+n-m,\D_3)\,\frac{\zb_{13} \bz_{23}}{z_{13} z_{23}}\,\Bigg]~~.
    \end{aligned}$}%
\end{equation}
Note the single-valued $\eta$-dependence. This result is a slight extension of the consecutive collinear limits considered in \cite{Ebert:2020nqf}.

\section{\texorpdfstring{Special Case: $\phi^3$ Theory}{Special Case: phi3 Theory}}\label{sec:phi3}

In this section, we will look into an example where the double residue condition is violated --- the celestial massless $\phi^3$ theory --- and show that the obstruction for hard celestial amplitudes comes from the existence of branch cuts in the celestial multi-OPEs even at tree level. Unlike pure Yang-Mills and pure gravity discussed in the previous section, in $\phi^3$ theory, the $1/s_{123}$ propagator appears manifestly in the 3-collinear splitting function. 
As we will explicitly see shortly, the appearance of $1/s_{123}$ is responsible for the branch cuts observed in the celestial multi-OPEs and also leads to the discovery of a new term in the celestial OPE that we will discuss in the next section.

Our procedure is as follows. We start with the holomorphic 3-collinear splitting function in momentum space and then implement the Mellin transform to obtain the celestial 3-OPE for $\phi^3$ theory.
The main part of this section will focus on the analytical computation of the integral coming from the Mellin transform and its behavior near special points.

\paragraph{Holomorphic 3-collinear limit} First, we adopt the same parametrization for the holomorphic 3-collinear limit as in section~\ref{sec:mom-space-multicollinear}:
\be 
z_1 ~=~ z_3 + \eps ~~,~~~ z_2 ~=~ z_3 + \eta\,\eps ~~.
\ee
The 3-collinear splitting function for $\phi^3$ theory in momentum space is simply the multiplication of propagators. All three diagrams discussed in section~\ref{sec:mom-space-multicollinear} will contribute and therefore we have
\begin{equation}
    {\rm Split}[123\to P] ~=~ \lambda^2\,\left( \frac{1}{(p_1+p_2)^2} + \frac{1}{(p_1+p_3)^2} + \frac{1}{(p_2+p_3)^2} \right) \frac{1}{(p_1+p_2+p_3)^2}
\end{equation}
where $\lambda$ is the 3-point coupling. Plugging in the momentum parametrization in \eqref{eq:momparam}, the splitting function is
\begin{equation}
    \begin{split}
        {\rm Split}[123\to P] ~&=~ \frac{\lambda^2}{16}\, \frac{ \frac{1}{\omega_1\omega_2 z_{12} \zb_{12}} + \frac{1}{\omega_1\omega_3 z_{13} \zb_{13}} + \frac{1}{\omega_2\omega_3 z_{23} \zb_{23}} }{\omega_1\omega_2 z_{12} \zb_{12} + \omega_1\omega_3 z_{13} \zb_{13} + \omega_2\omega_3 z_{23} \zb_{23}} ~~.
    \end{split}
\end{equation}
It is known, for example by noting the nonzero angle-bracket-weight $-1$ part of the 4-point amplitude~\cite{Ball:2022bgg}, that amplitudes in $\phi^3$ theory fail the double residue condition, so as a consistency check we compute the double residue condition of our splitting function. We find
\be \ba
\res{z_2\to z_3} \, \res{z_1\to z_2} {\rm Split}[123\to P] ~& =~
\frac{\lambda^2}{16\,\omega_1\omega_2\zb_{12}} \frac{1}{\omega_1\omega_3\zb_{13} + \omega_2\omega_3\zb_{23}} ~~,\\
\res{z_1\to z_3} \, \res{z_2\to z_3} {\rm Split}[123\to P] ~&=~
\frac{\lambda^2}{16\,\omega_2\omega_3\zb_{23}} \frac{1}{\omega_1\omega_2\zb_{12} + \omega_1\omega_3\zb_{13}}  ~~,\\
\res{z_2\to z_3} \, \res{z_1\to z_3} {\rm Split}[123\to P] ~& =~
\frac{\lambda^2}{16\,\omega_1\omega_3\zb_{13}} \frac{1}{-\omega_1\omega_2\zb_{12} + \omega_2\omega_3\zb_{23}}  ~~.
\ea \ee
From these we can see that the double residue condition is not satisfied, and moreover that the $1/s_{123}$ propagator is the key player in violating the double residue condition. 
For the purpose of simplification, we adopt the following parametrization for $\bz_i$ variables,
\begin{equation}
\bz_1 ~=~ \bz_3 + \bar{\eps} ~~,~~~ \bz_2 ~=~ \bz_3 + \bar{\eta}\bar{\eps} ~~,
\end{equation}
and note that we are considering the holomorphic collinear limit, therefore $\bar{\eps}$ and $\bar{\eta}$ are generic. Then, the 3-point splitting function becomes
\begin{equation}
    {\rm Split}[123\to P] ~=~ \frac{\lambda^2}{16\,\varepsilon^2\bar{\eps}^2}\, \frac{\frac{1}{\omega_1\omega_2 (1-\eta) (1-\bar{\eta})} + \frac{1}{\omega_1\omega_3 } + \frac{1}{\omega_2\omega_3 \eta \bar{\eta}}}{\omega_1\omega_2 (1-\eta) (1-\bar{\eta}) + \omega_1\omega_3  + \omega_2\omega_3 \eta \bar{\eta}}~~.
\end{equation}

\paragraph{Implement Mellin transform}
Following the same procedure as in section~\ref{sec:celestial-mulope}, direct computation yields
\begin{equation}
    \begin{split}
       & {\cal O}_{\D_1}(z_1,\bz_1) {\cal O}_{\D_2}(z_2,\bz_2) {\cal O}_{\D_3}(z_3,\bz_3)
        ~\sim~ \sum_{m=0}^\infty \,\frac{{\cal I}_m}{m!} \p_{\zb_3}^m\,\oh_{\Delta_1+\Delta_2+\Delta_3-4}(z_3, \zb_3)
    \end{split}
    \label{equ:phi3-3-OPE}
\end{equation}
with
\be \resizebox{0.9\textwidth}{!}{$%
{\cal I}_m ~=~ \scalemath{1.0}{\frac{\lambda^2 \bar\eps^{m-2}}{16\,\eps^2}} \scalemath{.95}{\int_0^1 \hspace{-1mm} ds\int_0^{1-s} \hspace{-2mm} dt \,  \frac{s^{\Delta_1-1} t^{\Delta_2-1} (1-s-t)^{\Delta_3-1}\left(\frac{1}{st (1-\eta)(1-\bar\eta)} + \frac{1}{s(1-s-t)} + \frac{1}{t(1-s-t) \eta \bar\eta}\right) (s + \bar\eta t)^m}{st (1-\eta)(1-\bar\eta) + s(1-s-t) + t(1-s-t) \eta\bar\eta}}~. $}%
\ee
We find that these integrals ${\cal I}_m$ have singular points at $\eta = 0,1,\infty$, corresponding to the consecutive collinear limits, and that there are branch cuts stretching between these points.

\paragraph{Master Integral}
If we binomially expand $(s+\bar\eta t)^m$ then all terms for all $m$ can be handled in terms of the master integral
\be
{\cal I}(\eta) ~\equiv~ \int_0^1 ds\,\int_0^{1-s} dt \frac{s^{\alpha_1} t^{\alpha_2} (1-s-t)^{\alpha_3}}{st (1-\eta)(1-\bar\eta) + s(1-s-t) + t(1-s-t) \eta\bar\eta} ~~,
\label{equ:master-integral}
\ee
which we view as a function of $\eta$. The integral is defined by analytic continuation from its region of absolute convergence. For generic $\eta, \bar\eta$, necessary conditions for absolute convergence are $\alpha_i > -1$ and $\alpha_i + \alpha_j > -1$ (for $i\ne j$) with $i,j = 1,2,3$. We assume these inequalities in the following discussion. Note that the technical difficulty to evaluate this integral comes from the denominator of the integrand which is quadratic in both $s$ and $t$. However, for our purpose, we are interested in the behaviors near $\eta\to 0$, $1$, and $\infty$. Under these limits, we are able to analytically compute the integral. Moreover, since the form of the integral is symmetric on these three limits, we will focus on $\eta\to 0$ behavior in what follows. The analysis for $\eta\to 1$ and $\eta\to \infty$ behaviors can be done in a similar manner.

\paragraph{Leading small-$\eta$ behavior of ${\cal I}(\eta)$}
First of all, we notice that when $\alpha_1 > 0$ the master integral simply converges at $\eta=0$, giving
\be {\cal I}(0) ~=~ B(\alpha_1, \alpha_2 + \alpha_3 + 1) B(\alpha_2 + 1, \alpha_3 + 1)\, {}_2F_1(1, \alpha_2 + 1, \alpha_2 + \alpha_3 + 2, \bar\eta) ~~.\ee
However, when $-1 < \alpha_1 < 0$, the integral converges generically but diverges near $\eta=0$ as $\eta^{\alpha_1}$. We can scale this out by changing variables as $s = \eta\sigma$, giving
\be {\cal I}(\eta) ~=~ \eta^{\alpha_1} \int_0^{1/\eta} d\s \int_0^{1-\eta\s} dt \frac{\s^{\alpha_1} t^{\alpha_2} (1 - \eta\s - t)^{\alpha_3}}{\s t(1-\eta)(1-\bar\eta) + \s(1 - \eta\s - t) + t(1 - \eta\s - t) \bar\eta}~~. \ee
One can show rigorously that the assumption $\alpha_1 < 0$ allows us to replace the upper integration bounds as $1/\eta \to \infty$ and $1-\eta\s \to 1$ in the $\eta\to 0$ limit. This gives
\be \ba &\lim_{\eta\to 0}\, \eta^{-\alpha_1}\, {\cal I}(\eta)  ~=~ \int_0^\infty d\sigma \int_0^1 dt\, \frac{\s^{\alpha_1} t^{\alpha_2} (1 - t)^{\alpha_3}}{\s t(1-\bar\eta) + \s(1 - t) + t(1 - t) \bar\eta} \\
~& =~ \frac{-\pi\,\bar\eta^{\alpha_1}}{\sin\pi\alpha_1} B(\alpha_1+\alpha_2+1, \alpha_1+\alpha_3+1) \, {}_2F_1(\alpha_1+1, \alpha_1+\alpha_2+1, 2\alpha_1+\alpha_2+\alpha_3+2, \bar\eta)~~. \ea \ee
Altogether, we get the overall leading small-$\eta$ behavior
\be \label{eq:asymp} 
{\cal I}(\eta) ~\sim~ C_1 ~+~ \eta^{\alpha_1} C_2 ~~,
\ee
where the coefficients are
\begin{equation}
    \begin{split}
        C_1 ~&=~ B(\alpha_1, \alpha_2 + \alpha_3 + 1) B(\alpha_2 + 1, \alpha_3 + 1)\, {}_2F_1(1, \alpha_2 + 1, \alpha_2 + \alpha_3 + 2, \bar\eta)~~,\\
        C_2 ~&=~ \frac{-\pi\,\bar{\eta}^{\alpha_1}}{\sin\pi\alpha_1}\, B(\alpha_1+\alpha_2+1, \alpha_1+\alpha_3+1) {}_2F_1(\alpha_1+1, \alpha_1+\alpha_2+1, 2\alpha_1+\alpha_2+\alpha_3+2, \bar\eta) ~~.
    \end{split}
\end{equation}
The terms exchange dominance depending on the sign of $\alpha_1$, but even when $C_1$ is dominant the branch cut coming from $\eta^{\alpha_1} C_2$ still provides the mechanism of obstruction to the double residue condition for hard celestial amplitudes.

\paragraph{Celestial multi-OPE for $\phi^3$}
Now we plug the leading small-$\eta$ behavior of the integral ${\cal I}(\eta)$ into the 3-OPE (\ref{equ:phi3-3-OPE}) and further simplify the result by assuming small $\bar\eps, \bar{\eta}$. Then the 3-OPE for $\phi^3$ reads
\be \resizebox{0.9\textwidth}{!}{$%
\boxed{ 
\oh_{\Delta_1}(z_1,\bz_1) \oh_{\Delta_2}(z_2,\bz_2) \oh_{\Delta_3}(z_3,\bz_3) ~\sim~ \left( \frac{1}{z_{13} z_{23} \zb_{13} \zb_{23}} {\cal C}_1 + \frac{(z_{23} \zb_{23})^{\Delta_1-2}}{(z_{13} \zb_{13})^{\Delta_1}} {\cal C}_2 \right)  \oh_{\Delta_1+\Delta_2+\Delta_3-4}(z_3, \zb_3) 
} $}%
\label{equ:OOO-phi3-final}
\ee
where the coefficients are
\begin{equation}
    \begin{split}
         {\cal C}_1 ~&=~ \frac{\lambda^2}{16}B(\D_1-1,\D_2+\D_3-3)\,B(\D_2-1,\D_3-1) ~~,\\
         {\cal C}_2 ~&=~ \frac{-\lambda^2\pi}{16\sin\pi\D_1} B(\D_1+\D_2-2,\D_1+\D_3-2) {\left[ 1 + \frac{\D_1+\D_3-2}{\D_1+\D_2-3} + \frac{\D_1+\D_2-2}{\D_1+\D_3-3}  \right]}~~.
    \end{split}
\end{equation}
We emphasize that (\ref{equ:OOO-phi3-final}) is the result under the $\eta\to 0$ limit, which corresponds to $z_{23} \ll z_{13},z_{12}$. Moreover, while only one of the three diagrams in section~\ref{sec:mom-space-multicollinear} contributes to the first term in (\ref{equ:OOO-phi3-final}), which is the usual consecutive celestial OPE, all three diagrams contribute to the second term.

\section{Closing Remarks}\label{sec:conclusion}

In this work, we began by introducing the holomorphic 3-collinear limit, which differs in general from the true multicollinear limit that has already appeared in the amplitudes literature. We derived a general formula for the corresponding splitting function and showed that it simplifies in theories whose celestial currents satisfy the Jacobi identity such as Yang-Mills and pure gravity, reducing to a weighted sum over the consecutive 2-collinear limits. We then extracted the celestial 3-OPE by massaging the Mellin transform of the splitting function.  

After providing these general constructions, we narrowed our focus to $\phi^3$ theory as a tractable example that violates the double residue condition, but where we can still carry out explicit computations to elucidate the origin of these violations. 
In particular, we evaluated the leading celestial 3-OPE coefficient in the limit of particles $1$ to $3$ coming together, with $z_{23} \ll z_{13} \ll z_{1i}$ for $i\ge 4$, and found that it contains not only the term expected from consecutive application of the familiar celestial OPE but also a term with a branch cut. The presence of the branch cut explains how Jacobi can fail for hard celestial amplitudes, and does so in a way that respects locality: on the Euclidean locus where $z, \zb$ are complex conjugates the holomorphic and antiholomorphic branch cuts cancel out. In combination with locality, the branch cut term suggests that in the celestial OPE between two particles there should be an additional term with a branch cut:
\be 
\begin{split}
    \oh_{\D_2}(z_2,\bz_2) \oh_{\D_3}(z_3,\bz_3) ~\sim&~ \frac{\lambda}{4\,z_{23}\bz_{23}} B(\D_2-1,\D_3-1) \oh_{\Delta_2+\Delta_3-2}(z_3,\bz_3) \\
    &\qquad\qquad\qquad ~+~ \frac{{\cal C}}{(z_{23}\bz_{23})^{2-\Delta_1}} {\cal R}_{2\Delta_1+\Delta_2+\Delta_3-4}(z_3,\bz_3)~~. 
\end{split}
\ee
In the latter term we propose a new operator ${\cal R}_{\D}$ with scaling dimension $\D$, coming with some coefficient ${\cal C}$. This term is new in the celestial literature and even leading when ${\rm Re} \, \Delta_1 < 1$. However, locality further suggests that the OPE should not treat $\Delta_1$ as special since $\oh_{\D_1}$ is far away. 
We address this by broadening our OPE ansatz to allow for two continuous parameters $(\D,\s)$ in the unknown operator ${\cal R}_{\D}^{\sigma}$ and working in terms of abstract OPE coefficients $C_{IJ}{}^K$. We write
\be \begin{split}
    \oh_{\D_2}(z_2,\bz_2) \oh_{\D_3}(z_3,\bz_3) ~\sim&~ \int d\D\,\frac{C_{\D_2,\D_3}{}^{\D}}{(z_{23} \zb_{23})^{\half(\D_2 + \D_3 - \D)}} \,\oh_{\Delta}(z_3, \zb_3)\\
    &\qquad\qquad\qquad~+~ \int d\Delta \, d\sigma \, \frac{ C_{\D_2,\D_3}{}^{(\D,\sigma)}}{(z_{23} \zb_{23})^{\half(\D_2 + \D_3 - \D)}} \,\mathcal{R}^{\sigma}_\D(z_3, \zb_3)~~.
\end{split} \ee
Note that for the first term on the RHS, the leading singular term is the usual celestial OPE with coefficient
\begin{equation}
    C_{\D_2,\D_3}{}^{\D} ~=~ \frac{\lambda}{4}\,B(\D_2-1,\D_3-1)\,\d(\D-\D_2-\D_3+2)~~.
\end{equation}
The requisite OPE between $\oh_{\Delta_1}$ and ${\cal R}_\Delta^\s$ is then
\begin{equation}
    \oh_{\D_1}(z_1,\bz_1) \, {\cal R}^{\s}_{\D}(z_3,\bz_3) ~\sim~ \int d\D'\frac{C_{\D_1,(\D,\s)}{}^{\D'}}{(z_{13}\zb_{13})^{\half(\D_1+\D-\D')}}\,\oh_{\D'}(z_3, \zb_3)~~.
\end{equation}
In this language \eqref{equ:OOO-phi3-final} reads
\be \oh_{\D_1} \oh_{\D_2} \oh_{\D_3} ~\sim~ (\dots) ~+~ \int \frac{d\D' \, d\D \, d\sigma\,C_{\D_2,\D_3}{}^{(\D,\sigma)} \, C_{\D_1,(\D,\sigma)}{}^{\D'}}{(z_{23}\zb_{23})^{\half(\D_2+\D_3-\D)} (z_{13}\zb_{13})^{\half(\D_1+\D-\D')}} \,\oh_{\D'}(z_3, \zb_3)~~, \ee
where
\begin{equation}
   \int d\s \, C_{\D_2,\D_3}{}^{(\D,\sigma)} \, C_{\D_1,(\D,\sigma)}{}^{\D'} ~=~ {\cal C}_2\,\d(\D'-\D_1-\D_2-\D_3+4)\,\d(\D-2\D_1-\D_2-\D_3+4)~~.
\end{equation}
While the individual OPE coefficients are unknown to us, they are significantly constrained by this concrete formula. The rest of our discussion is more speculative, but we can make an educated guess as to the identity of the operator ${\cal R}_\D^\s$. Like ${\cal R}_\D^\s$, two-particle operators have two continuous labels, one being the scaling dimension. Generally, the spin-0 operators constructed from two scalar fields can be defined as follows~\cite{YSL,kp}
\begin{equation} \label{eq:2partop}
    :\oh^{(\rho)}\oh:_{\D}\hspace{-.5mm}(z,\bz) ~\equiv~ \int_0^{\infty}d\o\,\o^{\D-\rho-1}\,\int_0^{\o}\,d\o_1\,\o_1^{\rho-1}\,a^{\dagger}(\o_1,z,\bz)a^{\dagger}(\o-\o_1,z,\bz) ~~,
\end{equation}
where $\D$ labels the total scaling dimension. The fact that there are two continuous labels is consistent with the fact that a two-particle operator located at a single point on the celestial sphere is labeled by two energies $\omega_i$, or equivalently null times $u_i$. (Analogously, the single-particle operator has one continuous label $\D$, $u$, or $\omega$ depending on your choice of basis~\cite{Donnay:2022sdg}). The particular combination in \eqref{eq:2partop} is then adapted to the boost eigenbasis.

What is interesting about CCFT is that it is nominally organizing scattering in terms of the full OPE in the collinear limit which is different than the perturbative expansion. However, if we view these objects in terms of their uplift to 4D correlation functions of operators at null infinity, we see that up to order $\lambda^2$ only the two-particle operators would appear in the connected correlation functions with two single-particle operators in $\phi^3$ theory at tree level. We can thus conclude that the new term ${\cal R}_{\D}^{\s}$ is a two-particle operator. Decomposing it as in (\ref{eq:2partop}) involves the multi-particle generalization of the RSW inner product of \cite{Cotler:2023qwh}. See~\cite{kp} for upcoming work on cataloging the multi-particle operators and~\cite{YSL} on their role in the celestial symmetry algebras. These are important steps towards applying more canonical bootstrap machinery to the celestial hologram.

\section*{Acknowledgements}

We would like to thank Justin Kulp, Andrzej Pokraka, and Akshay Yelleshpur Srikant for useful conversations, and Freddy Cachazo for comments on the draft. The research of AB, YH, and SP is supported by the Celestial Holography Initiative at the Perimeter Institute for Theoretical Physics and the Simons Collaboration on Celestial Holography. Research at the Perimeter Institute is supported by the Government of Canada through the Department of Innovation, Science and Industry Canada and by the Province of Ontario through the Ministry of Colleges and Universities. AB was also supported by the US Department of Energy under contract DE-SC0010010 Task F and by Simons Investigator Award \#376208.

\appendix

\section{Converting the Double Residue Condition to $\eps, \eta$}
\label{app:dblres}

In this appendix, we change variables for the double residue condition from $z_1, z_2$ to $\eps, \eta$, acting on a hard amplitude in either momentum space or the celestial basis. Recall $z_{13} = \eps$ and $z_{23} = \eta\eps$. We restrict attention to functions of the form $\frac{1}{\eps^2} f(\eta)$, since the subleading terms in $\eps$ will not contribute to the double residue condition. We also assume that the behaviors of $f(\eta)$ near $\eta = 0, 1, \infty$ are at most respectively $\oh(\frac{1}{\eta})$, $\oh(\frac{1}{\eta-1})$, $\oh(\frac{1}{\eta})$. Besides these assumptions we allow for arbitrary branch cuts in $\eta$. Our starting point is then
\be \left( \res{z_2\shortrightarrow z_3} \, \res{z_1\shortrightarrow z_2} - \res{z_1\shortrightarrow z_3} \, \res{z_2\shortrightarrow z_3} + \res{z_2\shortrightarrow z_3} \, \res{z_1\shortrightarrow z_3} \right) \frac{1}{z_{13}^2} f(z_{23}/z_{13})~~. \ee
The residues can be rewritten as limits with a given quantity held fixed. We will use a notation where $\lim_{a\to b;c}$ means the limit as $a\to b$ with $c$ held fixed. This sort of limit is appropriate for a 
 function of two variables. Note that factors involving only $c$ can be pulled through the limit. We have
\be \ba \bigg( \res{z_2\shortrightarrow z_3} \, \res{z_1\shortrightarrow z_2} & - \res{z_1\shortrightarrow z_3} \, \res{z_2\shortrightarrow z_3} + \res{z_2\shortrightarrow z_3} \, \res{z_1\shortrightarrow z_3} \bigg) \frac{1}{\eps^2} f(\eta) \\
~& =~ \left( \lim_{z_{23}\to 0} z_{23} \lim_{z_{12}\to 0} z_{12} - \lim_{z_{13}\to 0} z_{13} \lim_{z_{23}\to 0} z_{23} + \lim_{z_{23}\to 0} z_{23} \lim_{z_{13}\to 0} z_{13} \right) \frac{1}{\eps^2} f(\eta) \\
~& =~ \left( \lim_{\eta\eps\to 0} \eta\eps \lim_{(1-\eta)\eps\to 0;\eta\eps} (1-\eta)\eps - \lim_{\eps\to 0} \eps \lim_{\eta\eps\to 0;\eps} \eta\eps + \lim_{\eta\eps\to 0} \eta\eps \lim_{\eps\to 0;\eta\eps} \eps \right) \frac{1}{\eps^2} f(\eta) \\
~& =~ \left( \lim_{\eta\eps\to 0} \lim_{(1-\eta)\eps\to 0;\eta\eps} (1-\eta)\eta - \lim_{\eps\to 0} \lim_{\eta\eps\to 0;\eps} \eta + \lim_{\eta\eps\to 0} \lim_{\eps\to 0;\eta\eps} \eta \right) f(\eta) ~~.\ea \ee
Note that the $\eps$ dependence has dropped out completely, that $\lim_{(1-\eta)\eps\to 0;\eta\eps}$ involves $\eta\to 1$, that $\lim_{\eta\eps\to 0;\eps}$ involves $\eta\to 0$, and that $\lim_{\eps\to 0;\eta\eps}$ involves $\eta\to\infty$. Then these three initial limits can be written simply as $\eta$ limits, and the three subsequent limits are trivial and can be dropped. This gives
\be \left( \res{z_2\shortrightarrow z_3} \, \res{z_1\shortrightarrow z_2} - \res{z_1\shortrightarrow z_3} \, \res{z_2\shortrightarrow z_3} + \res{z_2\shortrightarrow z_3} \, \res{z_1\shortrightarrow z_3} \right) \frac{1}{\eps^2} f(\eta) ~=~ \left( -\res{\eta\to 1} - \res{\eta\to 0} + \lim_{\eta\to\infty} \eta \right) f(\eta) ~~.\ee
Thus the double residue condition on a hard momentum space amplitude amounts to
\be 0 ~\stackrel{?}{=}~ \left( -\res{\eta\to 1} - \res{\eta\to 0} + \lim_{\eta\to\infty} \eta \right) {\rm Split}[1^{s_1}2^{s_2}3^{s_3}\to J^{s_J}] \ee
and similarly for the double residue condition on hard celestial amplitudes, with the splitting function replaced by the 3-OPE. Note that analytic $\eta$-dependence on the doubly-punctured plane $\mathbb{C} - \{0, 1\}$ would be sufficient to guarantee that this condition holds, as can be seen by contour pulling.

\bibliographystyle{utphys}
\bibliography{main}

\end{document}